\documentclass{elsart}
\usepackage[dvips]{graphicx}
\usepackage{epsfig}

\begin{document}
\begin{frontmatter}
\title{The Baryon Anomaly: Evidence for Color Transparency and Direct Hadron Production at RHIC }

\author{Stanley J. Brodsky} 
\address{Stanford Linear Accelerator Center, Stanford University, Stanford,
California 94309}

\author{Anne Sickles}
\address{Brookhaven National Laboratory, Upton, New York 11973}

\begin{abstract} 
We show that the QCD color transparency of higher-twist contributions to inclusive hadroproduction 
cross sections, where baryons are produced directly in a short-distance subprocess,  
can explain several remarkable features of high-$p_T$ baryon production in heavy ion collisions 
which have recently been observed at RHIC: (a) the anomalous increase of the proton-to-pion ratio 
with centrality (b): the increased power-law fall-off at fixed $x_T=2p_T/ \sqrt s$ of the charged 
particle production cross section in high centrality nuclear collisions, and (c): the anomalous 
decrease of the number of same-side hadrons produced in association with a proton trigger as the 
centrality increases.   We show that correlations between opposite-side hyperons and kaons
can provide a clear signature of higher-twist contributions. These phenomena emphasize the importance of understanding hadronization at the amplitude level in QCD illustrate how heavy 
ion collisions can provide sensitive 
tools for interpreting and testing fundamental properties of QCD.
\end{abstract}
\end{frontmatter}
\maketitle
\section{Introduction}
One of the most surprising results observed at RHIC is the behavior of the ratio of protons to 
pions produced at large transverse momenta in heavy ion collisions.  Intuitively, one would expect that protons and other baryons would be depleted relative to mesons as the overlap of the colliding nuclei is increased. However,
as shown in Fig.  \ref{figA1}, the $p/\pi$ and $\bar{p}/\pi$ ratio at $p_T \sim 4 $ GeV/c measured at RHIC increases with the centrality of the heavy ion collision.

The standard perturbative QCD approach to hadron production at large transverse momentum is based on 
elementary  leading-twist quark and gluon hard scattering processes followed by jet hadronization.
For $p_T>$2 GeV/c the $\pi^0$~\cite{Adare:2007dg} spectra in p+p collisions
at $\sqrt{s}$=200 GeV appear to be well described by next to leading
order pQCD calculations.  In Au+Au collisions the  produced high transverse momentum quarks and gluons must also traverse  a zone of hot nuclear
medium ~\cite{Arsene:2004fa,Back:2004je,Adams:2005dq,Adcox:2004mh}.
The scattered partons then lose energy
traversing the dense colored matter and
fragment into hadrons far from the collision region according to the same fragmentation 
process as in p+p collisions.  In this scenario the ratio of particles as a function
of $p_T$ should be nearly independent of the collision system.  
However, the data~\cite{Adler:2003kg,Abelev:2007ra}
show large modifications of the particle ratios for 2$<p_T<$6 GeV/c.

\begin{figure}[htb]
\centering
\includegraphics[angle=0,width=1.0\linewidth]{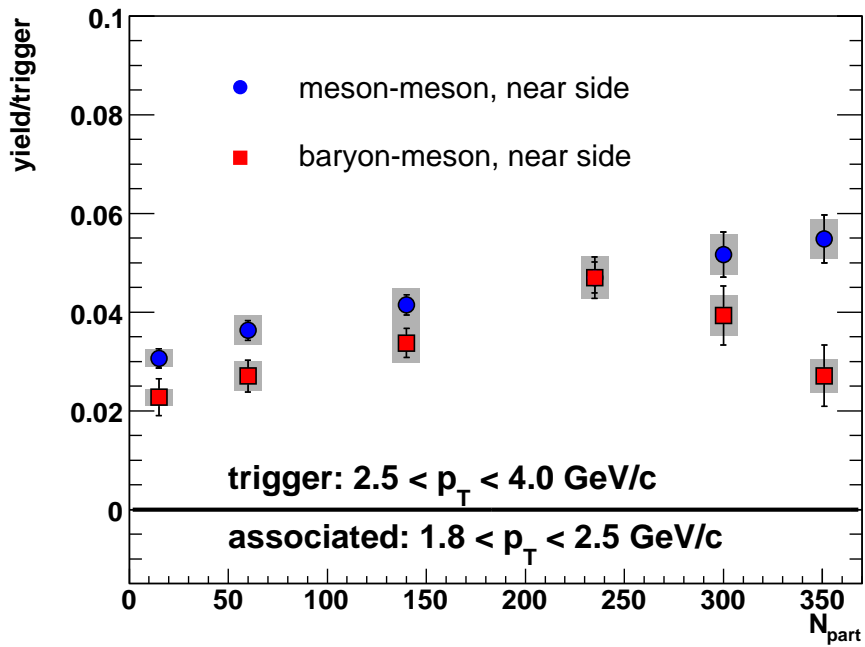}
\caption{ Same-side and away side correlated hadrons for meson and baryon triggers as a function of
$N_{part}$.  The number of same-side particles associated with a meson trigger
increases monotonically with the size of the collision system.
In contrast, he number of same-side particles associated with a proton trigger decreases  as $N_{part}$ increases.
From Ref.  \cite{Adare:2006nn}  } \label{figA2}
\end{figure}

Additional information on the anomalous baryon-to-meson particle ratios can be obtained by studying the correlations between the produced hadron and other hadrons produced on the same side at nearby rapidities~\cite{Adler:2004zd,Adare:2006nn}.
These correlations show an unexpected dependence on particle type.
The number of particles associated with a meson ($\pi^{\pm}$,$K^{\pm}$) 
at 2.5 $< p_T <$ 4.0 GeV/c increases linearly
with the number of nucleons from the incoming nuclei participating in the collision, $N_{part}$.  
This increase is qualitatively understood as the lost energy from the parton
producing additional hadrons correlated with the jet direction.
In contrast the number of particles associated with a high $p_T$ proton or anti-proton 
trigger decreases with $N_{part}$ in the most central collisions; see Fig. \ref{figA2}.  
This anomalous difference between the nuclear dependence of pion and proton production  
are inconsistent with the standard perturbative QCD picture
of hard scattering  followed by vacuum fragmentation.  

The increased baryon/meson ratios in Au+Au collisions have been explained by attributing
final state hadron formation for 2$<p_T<$6 GeV/c to quark 
coalescence~\cite{Hwa:2002tu,Greco:2003xt,Fries:2003kq}.  
In these models quarks (and anti-quarks) close in phase space
recombine to form the final state hadrons.  Such models favor baryons, with
three valence quarks, since the hadron momentum is the sum of the quark momenta.
However, these models have not been 
able to explain the centrality and particle type dependence of same side 
correlations~\cite{Fries:2004hd,Fries:2005is}.

We propose that the large baryon/meson ratios at RHIC collisions are the result
of baryons directly produced in the hard scattering through higher-twist subprocesses such as $q q \to B \bar q$. The
 baryon is initially produced in a color-singlet configuration; its transverse size is small reflecting the 
 high transverse momenta exchanged within the subprocess.
Higher-twist baryon and anti-baryon production should occur in both p+p and Au+Au collisions.  
However, in Au+Au collisions the small size of the baryons should enable them
to traverse the matter with minimal interactions as predicted by color transparency.
The medium in Au+Au collisions 
then acts as a filter;
partons from leading-twist scattering are suppressed by losing energy and those
produced in direct processes remain.

Higher-twist semi-exclusive subprocesses~\cite{Brodsky:1998sr} where hadrons interact directly in the hard subprocess are a natural feature of QCD.  Multi-parton and semi-exclusive subprocesses underly the analysis of hard exclusive processes such as deeply virtual Compton scattering, deeply virtual meson production, fixed-angle scattering, and elastic and inelastic form factors at large momentum transfer.  A particularly important example for inclusive reactions is the Drell-Yan process $\pi p \to \gamma^* X$ where the direct $n_{active =5} $ higher-twist subprocess $\pi q \to \gamma^* q$ dominates lepton pair production at high $x_F, $  explaining the constant behavior of the cross section as a function of the  parton momentum fraction and the observed dominance of longitudinally polarized virtual photons~\cite{Berger:1979du}.   The higher-twist amplitude is decreased by a factor of $f_\pi / Q$ relative to the leading-twist amplitude; however this is over-compensated by the lack of phase space suppression of the direct process at $x_F \to 1$.  The non-perturbative wavefunction which controls  the direct higher-twist process  $\pi q \to \gamma^* q$  is the gauge invariant and frame-independent pion distribution amplitude~\cite{Lepage:1980fj} $\phi_\pi(x)$.  The shape and normalization of hadronic distribution amplitudes can now be predicted  using the AdS/QCD correspondence~\cite{Brodsky:2008pg}.

\section{Scaling Behavior of Hard Hadron Production in QCD}
\begin{figure}[htb]
\centering
\includegraphics[angle=0,width=1.0\linewidth]{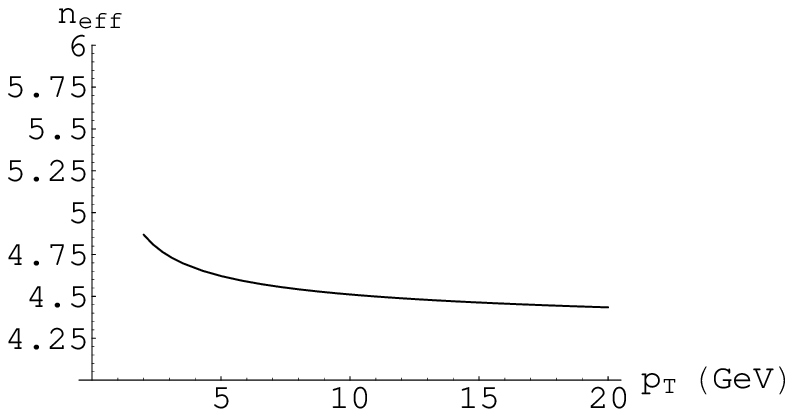}
\caption{Modification of scale-invariance from the logarithmic running of the QCD coupling constant and DGLAP evolution at $x_T=$0.03. From Ref. \cite{Brodsky:2005fza}
} \label{figA4}
\end{figure}

The most important discriminant of the twist of a pQCD  subprocess in a hard hadronic  collision is the scaling of the inclusive cross section
$${d\sigma\over d^3 p/E}(p p  \to H X)  = {F(x_T, \theta_{cm})\over p_T^n}$$ 
at fixed $x_T = {2p_T/ \sqrt s} $ and $\theta_{cm}$
In the original parton model~\cite{Berman:1971xz} the power fall-off is simply  $n= 4$ since the underlying $q q \to qq$ subprocess amplitude for point-like partons is scale invariant, and there is no dimensionful parameter in the theory. The Bjorken scaling of the deep inelastic lepton cross section $\ell p \to \ell^\prime X$  is based on the same scale-invariance principle.  In a full perturbative QCD analysis based on 2-to-2 quark and gluon subprocesses, the scale-invariance of the inclusive cross section is broken by the logarithmic running of the running coupling and the evolution of the structure functions and fragmentation functions.  These effects increase the prediction 
for $n$ to $n = 4.5 \to 5$ as illustrated in Fig. \ref{figA4} ~\cite{Brodsky:2005fza}.

\begin{figure}[htb]
\centering
\includegraphics[angle=0,width=1.0\linewidth]{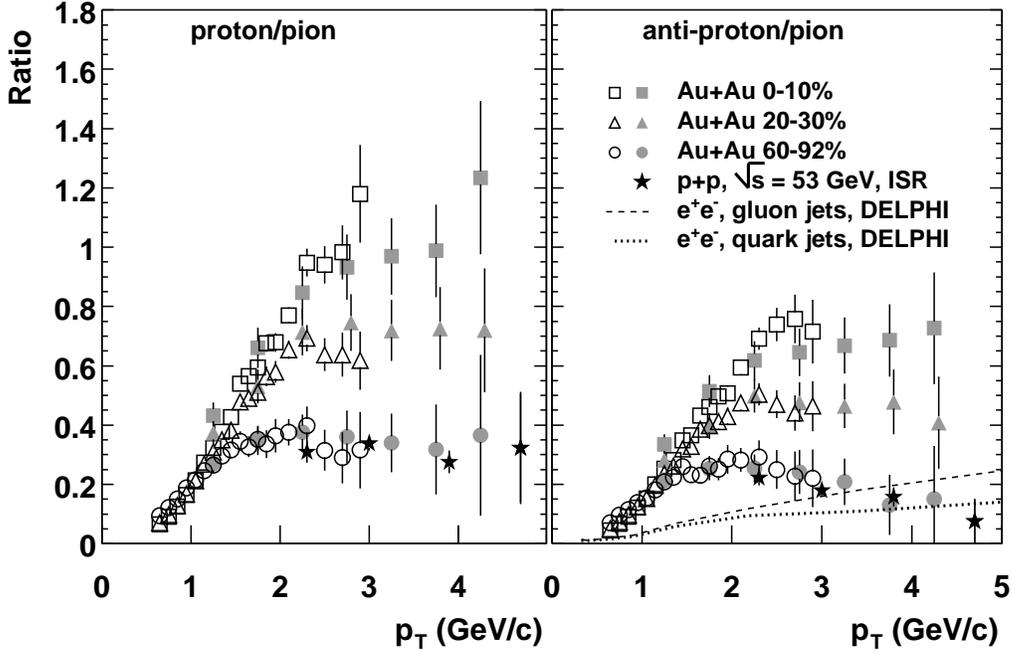}
\caption{Ratio of proton to pion and anti-proton to pion production as a function of $p_T$ for Au-Au collisions at at $\sqrt s = 200$ GeV for different centrality. The $p/\pi$ and 
$\bar{p}/\pi$ ratio
increase as collisions between the incoming nuclei become more central (the impact parameter
decreases).
  Open and filled symbols represent charged and neutral pions,  respectively. 
The stars show the particle ratio for $ p p$ collisions at $\sqrt s =53$ GeV.  The ratio for quark and gluon jet fragmentation are also shown.  From Ref. ~\cite{Adler:2003kg}} \label{figA1}
\end{figure}
%

%
%

%

\begin{figure}[htb]
\centering
\includegraphics[angle=0,width=1.0\linewidth]{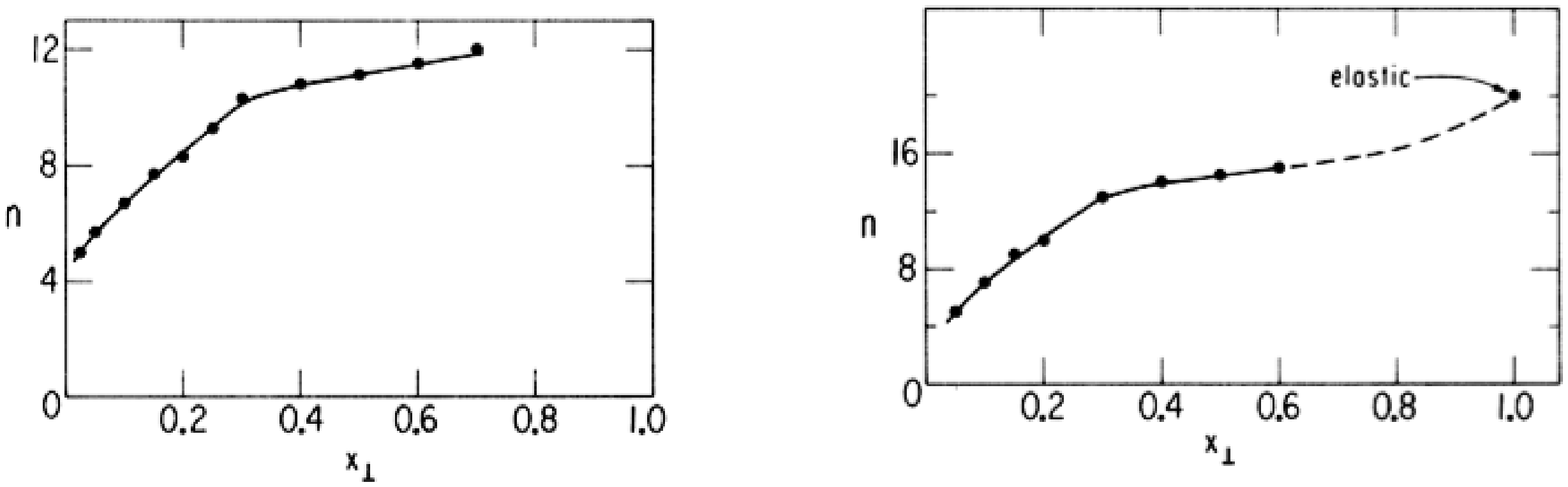}
\caption{ Summary of the measurements of the index $n$ controlling the effective 
power-law scaling of the inclusive cross section ${d\sigma\over d^3 p/E}(p p \to H X) = 
{F(x_T, \theta_{cm})\over p_{T}^n}$ at fixed $x_T = 2p_T/\sqrt{s}$ and $\theta_{cm}$
for pion (left panel) and  proton (right panel) hadroproduction, for $\sqrt{s}$ from 4.7-63GeV. 
From Ref. \cite{Cronin:1974zm}.
} \label{figA5}
\end{figure}

There have been extensive measurements of inclusive hadron production cross sections, particularly from the CERN ISR and  fixed-target experiments at Fermilab. 
The cross sections measured for $ p p \to \pi X$ and $p p \to p X$ are far 
from scale-invariant~\cite{Cronin:1974zm}; see Fig. \ref{figA5}. The power fall-off at fixed $x_T$ is consistent with the leading-twist pQCD prediction $n = 4.5 \to 5$  
only at the very smallest values of $x_T$.  Approximate scaling with $n=6.3$ is
observed for ISR through Tevatron energies for $x_{T}<$0.1~\cite{Adler:2003au}.
At higher $x_T$
it is observed to be a monotonically increasing function of $x_T$, reaching $n=20$ 
for $ p p \to p X$ at the exclusive limit $x_T \to 1.$  

\begin{figure}[htb]
\centering
\includegraphics[angle=0,width=1.0\linewidth]{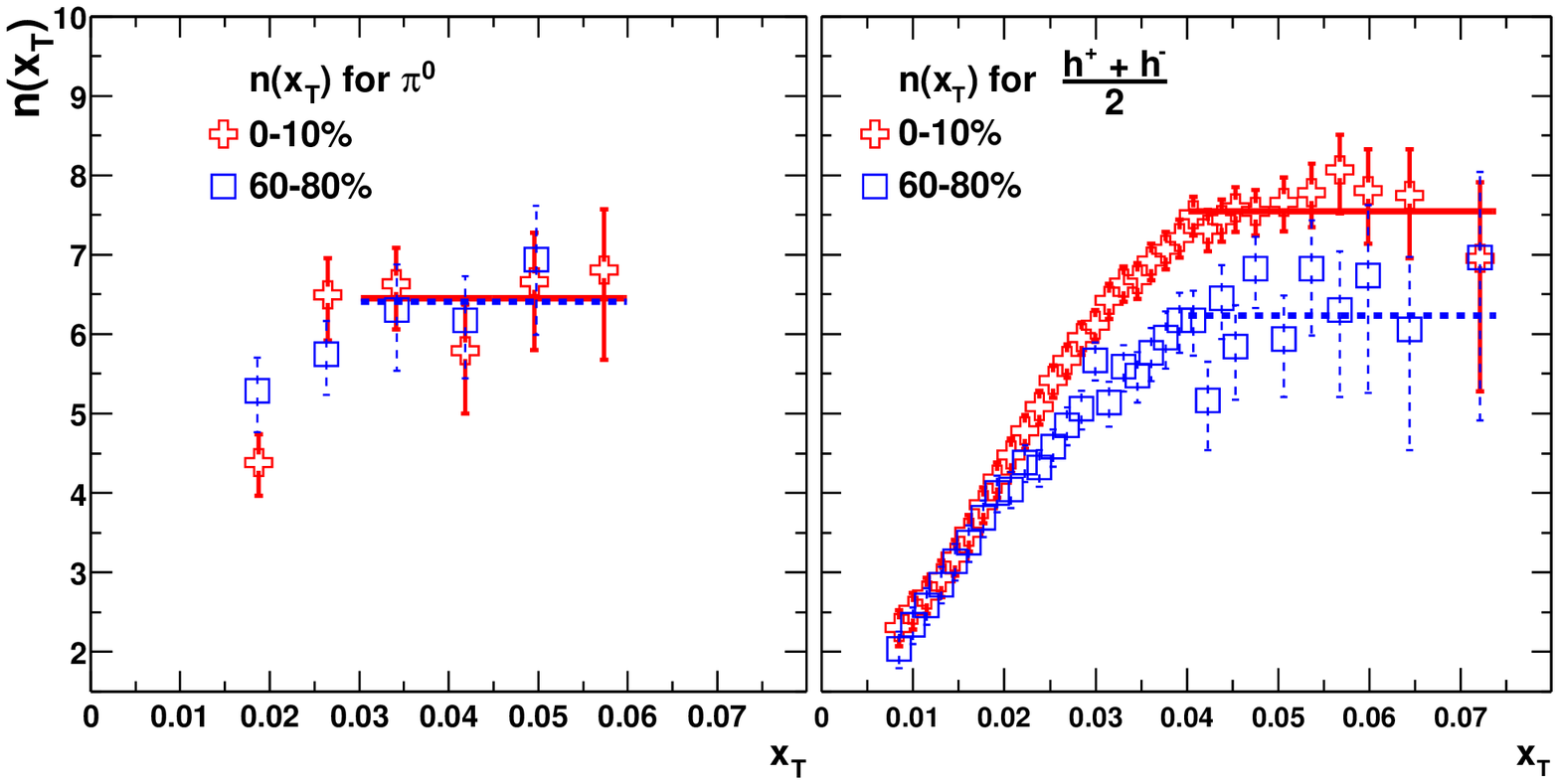}
\caption{ Effective power-law fall-off of the inclusive cross section for $\pi^0$ and charged particle hadroproduction at fixed $x_T$ and fixed $\theta_{cm}$ at RHIC energies. The power law increases as a function of $x_T$ and is different for central and peripheral collisions in the case of charged particle production.  The charged hadrons include protons and anti-protons.   From Ref. \cite{Adler:2003au}
} \label{figA6}
\end{figure}

In the case of RHIC,  the shape of the inclusive cross section for pion production measured in peripheral collisions at 
$\sqrt s = 200$ GeV~\cite{Adler:2003au}  appears to be in general agreement with NLO leading-twist QCD expectations~\cite{Adare:2007dg,Jager:2002xm}. 
However, as seen in Fig. \ref{figA6}, the scaling of the pion data at fixed $x_T$ for $0.03 < x_T < 0.06$ shows a 
rising behavior of $n(x_T)$ with an average value $n \sim 6.4\pm 0.5$~ \cite{Adler:2003au}.   Fig.  \ref{figA1} 
also shows  that the proton-to-pion and anti-proton to meson ratios measured in peripheral and central heavy ion 
collisions  differ from that of quark and gluon jets in $e^+ e^-$ annihilation~ \cite{Adler:2003kg}.  This 
breakdown of factorization also suggests that a description of the heavy ion hadroproduction data based solely on 
leading-twist contributions  is not adequate.   In contrast, the photon production 
cross section~\cite{Tannenbaum:2006ku,Adler:2006yt}
$ p p \to \gamma X$ at fixed $x_T$ scales over a large range of energies with a constant power $n \sim 5$  at $x_T < 0.04$, 
consistent  with the leading-twist pQCD prediction based on the $g q \to q \gamma$ subprocess. 
The  direct comparison 
of the $\gamma/ \pi$ ratio with theory at fixed $x_T$ would be illuminating;  if the leading-twist description is correct, the ratio should be nearly scale-invariant except for small corrections from jet fragmentation and the running coupling. The choice of renormalization scale for each subprocess, including the non-Abelian couplings, can be fixed using the 
BLM method~\cite{Brodsky:1982gc,Binger:2006sj},  thus eliminating one source of ambiguity of the leading-twist predictions. 

\begin{figure}[htb]
\centering
\includegraphics[angle=0,width=1.0\linewidth]{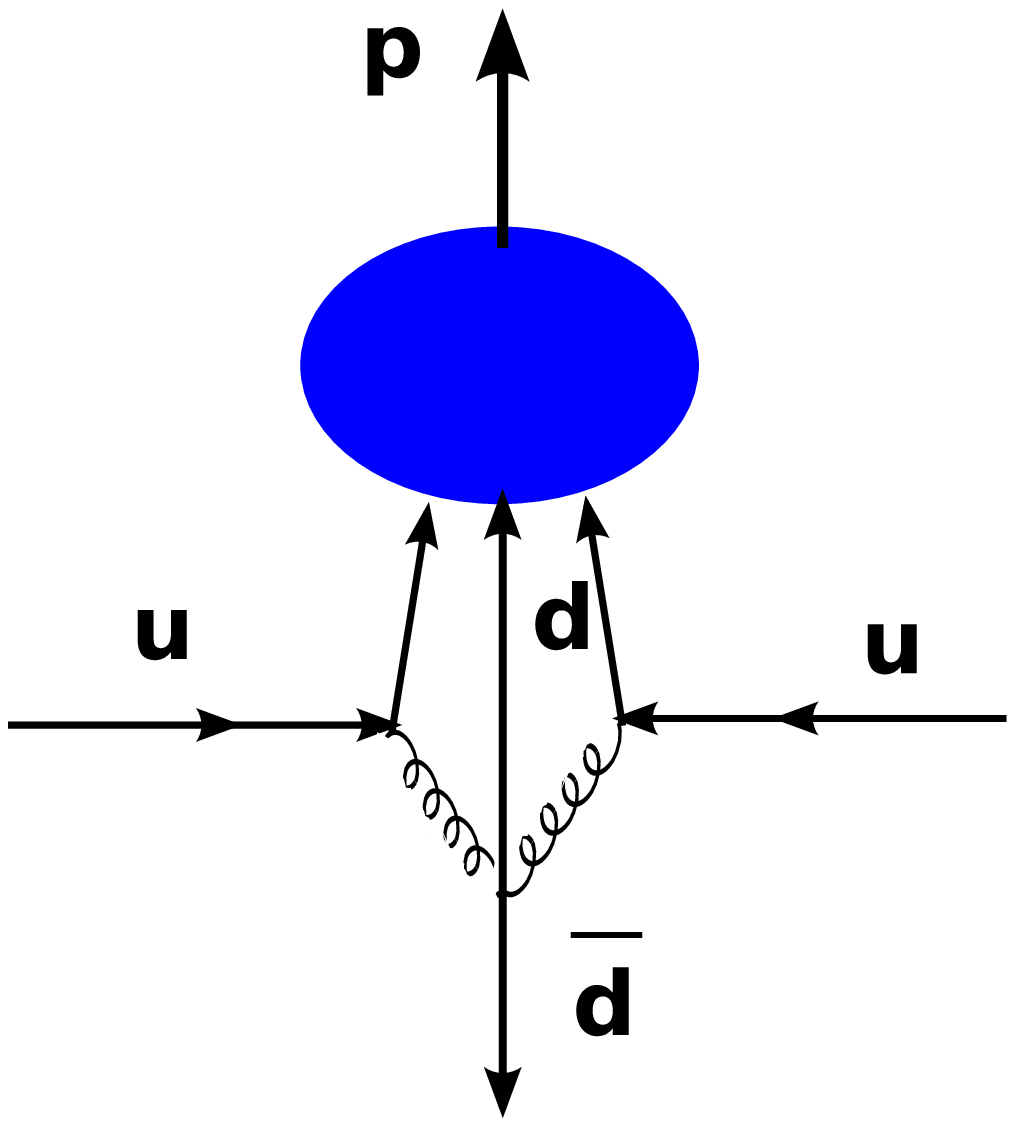}
\caption{ Higher-twist contribution to proton production at high $p_T.$  The proton is produced directly within the 6-parton hard subprocess. 
} \label{figA8}
\end{figure}

The seemingly anomalous scale-breaking behavior for hadroproduction can be naturally explained if in addition to the
 leading-twist processes, there are also contributions from  ``higher-twist" (multi-parton) processes.  As $x_T$ 
increases, it becomes more advantageous to produce the trigger hadron directly in a semi-exclusive hard 
subprocess~\cite{Brodsky:1998sr} such as $ g q \to \pi q$ or $ q q \to p \bar q$, since this avoids any waste of energy from jet fragmentation~\cite{Blankenbecler:1975ct}.  An example is illustrated in  Fig. \ref{figA8}. It  is also more energy efficient to scatter more than one  parton in the projectile, such as $q + (qq) \to q (qq) $ followed by fragmentation of the diquark to the trigger proton. In each case the penalty of the extra fall-off in $p_T$ from hadron compositeness or the diquark correlation scale is compensated by a lesser fall-off in $x_T$.

Dimensional counting rules provide a simple rule-of-thumb guide for the power-law fall-off of the inclusive cross section in both $p_T$ and $(1-x_T)$  due to a given subprocess ~\cite{Brodsky:1994kg}:
$$E{d\sigma\over d^3 p}(A B   \to C X) 
   \propto {(1-x_T)^{2 n_{spectator} -1}\over {p_T}^{2n_{active -4}}}$$ 
where $n_{active}$ is the ``twist", i.e., the number of elementary fields participating in the hard subprocess, and $n_{spectator}$ is the total number of constituents in $A, B$ and $C$ not participating in the hard-scattering subprocess. 
For example, consider $p p \to p X$.  
The leading-twist contribution from $q q \to q q$ has $n_{active} =4 $ and $n_{spectator}= 6.$ 
The  higher-twist subprocess $q q \to p \bar q$ has $n_{active} = 6 $ and $n_{spectator} = 4$ . This simplified model provides two distinct contributions to the inclusive cross section 
\begin{equation}
{d\sigma\over d^3 p/E}(p p   \to p X)  = A{(1-x_T)^{11}\over p_T^4} + B {(1-x_T)^{7}\over p_T^8} 
\label{eqspectra}
\end{equation}
and $n =  n(x_T)$ increases from 4 to 8 at large $x_T.$

In a general QCD analysis of inclusive hadroproduction one needs to sum over  all contributing leading and  higher-twist hard subprocesses. At $x_T=1$  the quarks in the protons must all scatter in an $n_{active}=12, n= 20, n_{spectator}=0$ exclusive subprocess.  In each case the nominal fall-off given by counting rules will be increased by the running of the QCD coupling and either  DGLAP evolution of the structure functions or ERBL evolution~\cite{Lepage:1980fj,Efremov:1979qk} of the distribution amplitudes for the directly-interacting hadrons. Although large $p_T$ hadron production at RHIC  is most likely dominated by leading-twist QCD processes~\cite{deFlorian:2007ty}, higher-twist subprocesses can play a significant role, particularly in the case of proton production.

\section{Color Transparency}

In higher-twist subprocesses such as $ g q \to \pi q$, $\pi q \to \gamma^* q$ or $q q \to p \bar q,$  
the wavefunction of a hadron enters directly into the amplitude. The dominant contribution comes from 
fluctuations of the  hadronic wavefunction where the quarks in the valence Fock state are at small 
impact separation  $b_\perp \sim 1/p_T$.  Interactions  with the external system are thus suppressed 
unless  the wavelength of the  exchanged gluon  is comparable to the transverse size of the color 
singlet system; i.e. $k_\perp \sim p_\perp$
The small-size color-singlet configurations of the hadron  can thus propagate through the nuclear 
medium with minimal hadronic interactions; i.e. they are {\it color transparent}~\cite{Bertsch:1981py}.

Color transparency~\cite{Brodsky:1988xz,Frankfurt:1992dx} is a fundamental property of QCD as a gauge 
theory of hadronic interactions. A clear empirical  demonstration has been given in diffractive di-jet 
production $ \pi A \to {\rm jet jet } A^\prime$  by the E791 experiment at Fermilab~\cite{Aitala:2000hc}; 
the forward amplitude for the diffractive production of high transverse momentum di-jets is found to scale 
as $A^\alpha$ where $\alpha \simeq 1$; i.e. the diffractive di-jet production amplitude is coherent on 
every nucleon in the nucleus. This is in dramatic contradiction to traditional Glauber theory where only 
nucleons on the periphery of the nucleus are effective.  Color transparency predictions for  quasi-elastic 
pion electoproduction $e A \to e^\prime  \pi^+ X$ have recently been verified at Jefferson 
Laboratory~\cite{:2007gqa}.


Color transparency provides an appealing explanation of the anomalous baryon to meson ratios
 observed at RHIC. For 
simplicity,  let us assume the two-component model for $p p \to p X$ given in Equation~\ref{eqspectra}.
The higher-twist term due to $q q \to p \bar q$ produces an isolated proton as a small color singlet 
which is unaffected by the nuclear environment;   in contrast, the leading-twist term
produces a high $p_T$ parton which propagates through the matter and loses energy.
The fragmentation process further suppresses the leading-twist term because the $p_T$ carried
by the parton is spread between many hadrons.
The increased relative contribution of higher-twist baryon production leads to the higher
effective power $n$ for charged hadrons seen in Fig.~\ref{figA6}.

Furthermore, since the increased importance of higher-twist contributions 
to the proton and anti-proton 
production cross section in highly central events ($N_{part} > 250$), we 
can also understand why the number of same side hadrons correlated with a the baryon trigger 
decreases (Fig. \ref{figA2}). 
The directly produced  proton interacts much less in the nuclear medium than a proton produced 
via jet fragmentation.   In contrast, the meson trigger  does not show this effect;
 the number of same-side mesons increase monotonically with $N_{part}$. 

Direct $p$ and $\bar{p}$ production could also explain higher $p_T$ ($p_T>$6 GeV/c) 
measurements at RHIC~\cite{mohantyqm08}.
Expectations from pQCD are that gluons passing through colored matter will lose more energy
than quarks because of the larger color factor; energy loss for a gluon should be
9/4 larger than for a quark.  At RHIC over 90\% of $p$ and $\bar{p}$
produced via leading-twist scattering at 6$<p_T<$12 GeV/c are
expected to come from gluon fragmentation. In contrast, approximately half of the pions are expected
from gluon fragmentation\cite{Ruan:2007gu,Albino:2005me}.  Because of the dominance
of gluon jets, $p$ and $\bar{p}$ spectra should 
be more suppressed than the pion spectra\cite{Wang:1998bha}.  The data, however,
show that even for 6$<p_T<$12 GeV/c
pions are more suppressed in central Au+Au collisions.
While the cross section for higher-twist processes falls off
with a higher power in $p_T$ than leading-twist processes, there could still be 
significant contributions from direct $p$ and $\bar{p}$ production.  
As seen in Equation \ref{eqspectra}, while the fall off of the higher-twist term is, 
in part, balanced by the lesser fall off in $x_T$.  

The $ p p \to \pi X$ cross section  also receives leading-twist fragmentation and direct higher-twist 
contributions from $g q\to  \pi q$, etc.; however, as seen from the power fall-off of $n(x_T)$ shown 
in Fig. \ref{figA5},  higher-twist  processes are evidently relatively  more significant for proton 
compared to pion triggers in the RHIC kinematic domain. Thus color transparency and direct hadron 
production is mostly associated with proton and other baryon triggers.


\section{Predictions for Baryons Containing $s$ and $\bar{s}$ Quarks}
Higher-twist processes are also expected to contribute to hyperon production. For example, 
a $\Lambda$ can be produced 
directly at large transverse momentum via the semi-exclusive subprocess 
$ u d \to \Lambda ~ \bar s$ in analogy to the $u u \to  p \bar d $ subprocess illustrated 
in Fig. \ref{figA8}.  In the case of $\Lambda$ production, the $s$ and $\bar{s}$ are
in opposing di-jets.  In contrast, in leading-twist parton scattering the strangeness of
particles in the opposing di-jets should be independent.  Measurements of correlations
between hyperons and charged kaons would be a clear signature of baryon production
via higher-twist subprocesses.  In that case, there should be an excess of $\Lambda$-$K^+$ 
and $\bar{\Lambda}$-$K^-$ correlations over $\Lambda$-$K^-$ and $\bar{\Lambda}$-$K^+$
correlations when the hyperon and kaon are in opposing di-jets.
The strength of this correlation should change with the relative contribution from
direct and fragmentation processes for $\Lambda$ production; measurements of the centrality
and $p^{\Lambda}_{T}$ dependence will constrain the contribution of direct processes to baryon 
production.
These correlations would not be naturally explained by quark coalescence models
and would provide a means of distinguishing between different baryon production mechanisms.

\section{Conclusions}

We have shown that the QCD color transparency of higher-twist contributions to inclusive hadroproduction 
cross sections, where baryons are produced directly in a short-distance subprocess,  
can explain several remarkable features of high-$p_T$ baryon production in heavy ion collisions 
which have recently been observed at RHIC: (a) the anomalous increase of the proton-to-pion ratio 
with centrality (b): the increased power-law fall-off at fixed $x_T=2p_T/ \sqrt s$ of the charged 
particle production cross section in high centrality nuclear collisions, and (c): the anomalous 
decrease of the number of same-side hadrons produced in association with a proton trigger as the 
centrality increases.   These phenomena emphasize the importance of understanding hadronization at the amplitude level in QCD. They also illustrate how heavy 
ion collisions can provide sensitive 
tools for interpreting and testing fundamental properties of QCD.

Clearly careful analyses and measurements at RHIC over a wide range of energies is needed to 
validate or disprove the importance of higher-twist direct reactions in hard ion collisions.   The scaling behavior of cross sections and particle ratios at fixed $x_T$ is the most direct measure of multiparton subprocesses.  We have also emphasized that strangeness correlations between opposite-side hyperons and kaons
can provide a clear signature of higher-twist contributions. Measurements of the associated particles in direct photon production 
will also be very valuable for understanding these remarkable features of QCD in the nuclear medium.

\smallskip
\noindent{\bf Acknowledgments:}\,\,

We thank David Morrison, Michael Tannenbaum, and Werner Vogelsang for helpful comments.
This work supported in part by the Department of Energy under contracts DE-AC02-76SF00515 (S.J.B) and DE-AC02-98CH1-886 (A.S.).
SLAC-PUB-13224.

\bibliography{baryon_paper}

\end{document}